\documentclass[aip,apl,a4,reprint,twoside,floatfix,superscriptaddress]{revtex4-1}
\usepackage{graphicx}
\usepackage{bm}

\begin{document}
\title[]{Observation of gyromagnetic reversal}
\author{Masaki Imai}
\thanks{These authors contributed equally to the work}
\affiliation{Advanced Science Research Center, Japan Atomic Energy Agency, Tokai 319-1195, Japan}
\author{Yudai Ogata}
\thanks{These authors contributed equally to the work}
\affiliation{Advanced Science Research Center, Japan Atomic Energy Agency, Tokai 319-1195, Japan}
\author{Hiroyuki Chudo}%
\affiliation{Advanced Science Research Center, Japan Atomic Energy Agency, Tokai 319-1195, Japan}
\author{Masao Ono}
\affiliation{Advanced Science Research Center, Japan Atomic Energy Agency, Tokai 319-1195, Japan}
\author{Kazuya Harii}
\affiliation{Advanced Science Research Center, Japan Atomic Energy Agency, Tokai 319-1195, Japan}
\author{Mamoru Matsuo}
\affiliation{Advanced Institute for Materials Research, Tohoku University, Sendai 980-8577, Japan}
\affiliation{Kavli Institute for Theoretical Sciences, University of Chinese Academy  of Sciences, Beijing, 100190, P.R.China}
\affiliation{Riken Center for Emergent Matter Science (CEMS), Wako 351-0198, Japan}
\author{Yuichi Ohnuma}
\affiliation{Advanced Science Research Center, Japan Atomic Energy Agency, Tokai 319-1195, Japan}
\affiliation{Kavli Institute for Theoretical Sciences, University of Chinese Academy  of Sciences, Beijing, 100190, P.R.China}
\affiliation{Riken Center for Emergent Matter Science (CEMS), Wako 351-0198, Japan}
\author{Sadamichi Maekawa}
\affiliation{Advanced Science Research Center, Japan Atomic Energy Agency, Tokai 319-1195, Japan}
\affiliation{Riken Center for Emergent Matter Science (CEMS), Wako 351-0198, Japan}
\affiliation{Kavli Institute for Theoretical Sciences, University of Chinese Academy  of Sciences, Beijing, 100190, P.R.China}
\author{Eiji Saitoh}
\affiliation{Advanced Science Research Center, Japan Atomic Energy Agency, Tokai 319-1195, Japan}%
\affiliation{Advanced Institute for Materials Research, Tohoku University, Sendai 980-8577, Japan}%
\affiliation{Institute for Materials Research, Tohoku University, Sendai 980-8577, Japan}%
\affiliation{Department of Applied Physics, The University of Tokyo, Hongo, Bunkyo-ku, Tokyo, 113-8656, Japan}

\date{\today}

\begin{abstract}
We report direct observation of gyromagnetic reversal, which is the sign change of gyromagnetic ratio in a ferrimagnet Ho$_3$Fe$_5$O$_{12}$, by using the Barnett effect measurement technique at low temperatures. 
The Barnett effect is a phenomenon in which magnetization is induced by mechanical rotation through the coupling between rotation and total angular momentum $J$ of electrons. The magnetization of Ho$_3$Fe$_5$O$_{12}$ induced by mechanical rotation disappears at 135~K and 240~K.
The temperatures correspond to the magnetization compensation temperature $T_{\rm M}$ and the angular momentum compensation temperature $T_{\rm A}$, respectively.
Between $T_{\rm M}$ and $T_{\rm A}$, the magnetization flips over to be parallel against the angular momentum due to the sign change of gyromagnetic ratio.
This study provides an unprecedented technique to explore the gyromagnetic properties.
\end{abstract}
\maketitle
Gyromagnetic ratio—the ratio of magnetic moment to angular momentum is an essential concept in magnetism.
An electron has negative gyromagnetic ratio, and its magnetic moment is antiparallel to the spin angular momentum.
In fact, most of magnets have a negative gyromagnetic ratio since their magnetization is caused by electrons.
In contrast, some magnets with multiple magnetic ions have a temperature dependent total gyromagnetic ratio, and even exhibit sign change: gyromagnetic reversal, which has manifested itself in some measurements\cite{Xin2006,Binder2006,Kim2017}.
Here, we report direct observation of gyromagnetic reversal in a ferrimagnetic insulator Ho$_3$Fe$_5$O$_{12}$ (HoIG).
We show that the net internal angular momentum of HoIG clearly disappears around 240 K, and, below the temperature, it flips over to be parallel to the magnetization.
This technique allowed us to investigate gyromagnetic properties, which is essential to accelerate spintronics and magnetic devices.

\begin{figure}
\includegraphics[width=\linewidth]{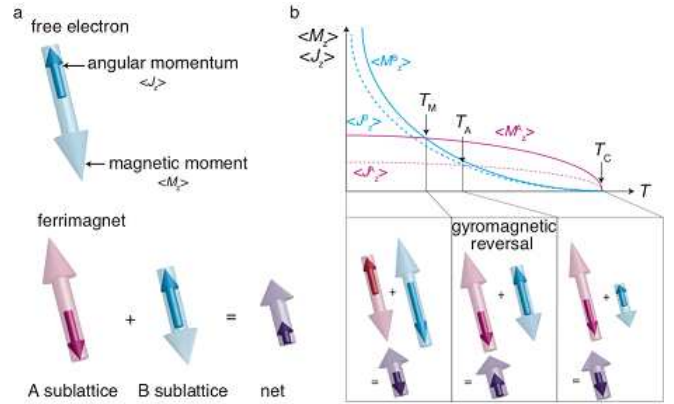}
\caption{Schematic illustration of angular momentum and magnetic moment.
{\bf a}: Relationship between angular momentum (dark blue arrow) and magnetic moment (light blue arrow) in a single electron.
Net angular momentum (dark purple arrow) and magnetization (light purple arrow).
Dark red (blue) arrow and light red (blue) arrow represent an angular momentum and magnetization in A (B) sub lattice.
{\bf b}: Temperature dependence of magnetization and angular momentum of A (B) sub lattice.
The temperatures $T_{\rm M}$, $T_{\rm A}$, and $T_{\rm C}$ represent magnetic compensation temperature, angular momentum compensation temperature, and Curie temperature, respectively. 
\label{fig1}}
\end{figure}

\begin{figure*}
\includegraphics[width=1\linewidth]{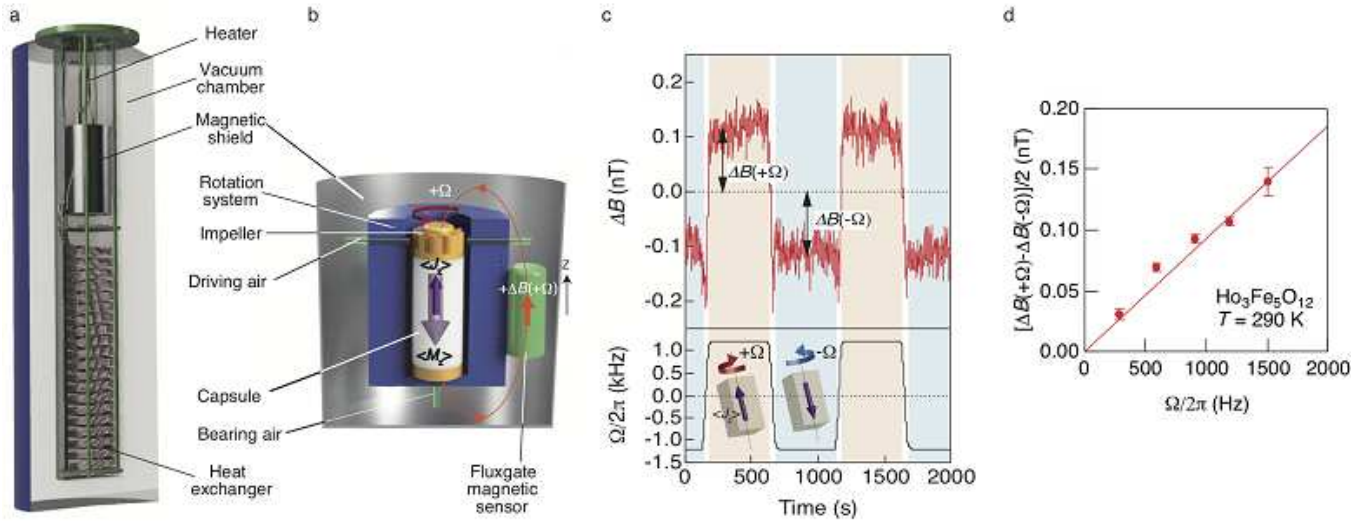}
\caption{Measurement of Barnett effect.
{\bf a}: Schematic illustration of the apparatus for measuring the Barnett effect.
{\bf b}: Directions of rotation, net angular momentum, magnetization, and the stray field $\Delta B$ are shown by arrows.
{\bf c}: Time evolution of rotation frequency and stray field $B$ of Ho$_3$Fe$_5$O$_{12}$ at $\Omega/2\pi$=1.2 kHz at 290K.
The Black and red lines show the rotation frequency and the stray field after offsetting the background, respectively. 
{\bf d}: Rotation frequency dependence of the stray field $\Delta B$.
\label{fig2}}
\end{figure*}

%%%%%%
In 1915, S. J. Barnett reported that matter is magnetized when it is rotated, and this effect has since been called the Barnett effect\cite{Barnett1915}.
His experiment demonstrated that the magnetization of a material carries internal angular momentum $J$ owing to electron spins\cite{Barnett1915,Barnett1935}.
On an electron, the angular momentum and the magnetic moment are antiparallel (Fig.\ref{fig1} {\bf a}), namely, the gyromagnetic ratio is negative.
Most magnets have thus internal angular momentum antiparallel to their magnetization.
When a magnet is rotated, the internal angular momentum is modulated via spin-rotation coupling, and then its magnetization is modified based on its gyromagnetic ratio.
This is the mechanism of the Barnett effect.
Therefore, the magnetization induced by the Barnett effect reflects the gyromagnetic ratio including its sign.

The gyromagnetic ratio shows various temperature dependence in some ferrimagnets.
In ferrimagnets,  magnetic moments in magnetic sublattices align antiparallel to each other(Fig.\ref{fig1} {\bf a}).  
In the case that the magnetization of two sublattices have different temperature dependence (Fig.\ref{fig1} {\bf b}) owing to differences in the intra-sublattice exchange interactions, these magnetizations can cancel each other at a certain temperature called the magnetization compensation temperature $T_{\rm M}$, at which the magnet has zero net magnetization.
In addition, the internal angular momentums can cancel each other at a different temperature called the angular momentum compensation temperature $T_{\rm A}$ in the case that magnetic ions on different sublattice have different $g$ factor (Fig.\ref{fig1} {\bf b}). 

Interestingly, across $T_{\rm M}$ or $T_{\rm A}$, the net magnetization reverses with respect to the angular momentum (Fig.\ref{fig1} {\bf b}): gyromagnetic reversal.
Between the two temperatures, the magnetization and the angular momentum are parallel.
The existence of $T_{\rm M}$ and $T_{\rm A}$ has often been discussed in terms of anomalies in various magnetic properties around the temperatures\cite{Ostler2012,Wangsness1953,Wangsness1954,Xin2006,Stanciu2006,Stanciu2007,Vahaplar2009,Radu2011,Kobayashi2005,Kim2017}.

In this letter, we demonstrated the gyromagnetic reversal using the apparatus possessing the capability to measure the Barnett effect at low temperatures.
We found that the magnetization of HoIG is reversed at two temperatures.
One coincides with $T_{\rm M}$ determined by the conventional magnetization measurement, and the other corresponds to $T_{\rm A}$.
In addition, the effective gyromagnetic ratio is positive in the region between $T_{\rm M}$ and $T_{\rm A}$, which is regarded as an additional state in terms of the gyromagnetic effect. 

\begin{figure*}
\includegraphics[width=0.7\linewidth]{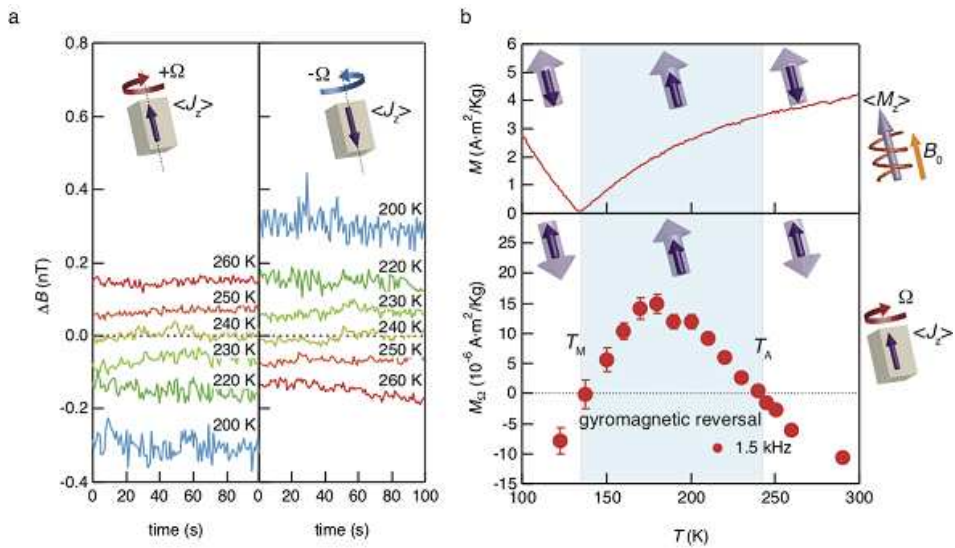}
\caption{Identification of gyromagnetic reversal region of Ho$_3$Fe$_5$O$_{12}$.
{\bf a}: Stray field in cases of positive and negative rotation at temperatures of 260 K to 200 K.
{\bf b}: The upper panel shows the temperature dependence of magnetization of Ho$_3$Fe$_5$O$_{12}$ in a magnetic field of 1000 Oe. 
The lower panel shows the temperature dependence of magnetization of Ho$_3$Fe$_5$O$_{12}$ due to mechanical rotation ($M_\Omega$) at a rotational frequency of $\Omega=1.5$~kHz (red solid circle). 
The error bars represent the standard deviation (1$\sigma$) estimated by fitting the data of Fig.3 {\bf a} to a line.
\label{fig3}}
\end{figure*}

Figure \ref{fig2}{\bf a} shows a schematic illustration of our experimental setup.
We used an air-driven rotor system\cite{Ono2015,Ogata2017A,Ogata2017B} and extended it to low-temperature measurement.
The rotation system was inserted into a cryostat, and the rotor was driven by nitrogen gas cooled through a heat exchanger.
The magnetization induced by the Barnett effect was estimated by measuring the stray field $\Delta B$ from the rotated sample, as shown in Fig.\ref{fig2}{\bf b}.
The value of $\Delta B$ was detected by using a fluxgate magnetic sensor placed near the sample.
The details of the apparatus are explained in the section\ref{sec:setup}.
As a sample, we used HoIG, which has two magnetic ions of Ho$^{3+}$ and Fe$^{3+}$ (see the section\ref{sec:sample}), because ferrimagnetic HoIG ($T_{\rm C}=567$~K) shows the magnetic compensation at $T_{\rm M}=137 \pm2$~K\cite{Pauthenet1958,Geller1965}, which is within the available temperature of our apparatus.

The time evolution of $\Delta B$ and the rotational speed $\Omega$ is shown in Fig. \ref{fig2}{\bf c}, indicates that the sample is magnetized along the direction of rotation because $\Delta B$ increases and decreases in sync with the switching of the direction of rotation.
By repeating the switching process, we averaged $\Delta B$ to be accurate.
The value of $\Delta B$ is stable during rotation and depends linearly on the rotational frequency, as shown in Fig.\ref{fig2}{\bf d}.
These results indicate that our custom-built measures the Barnett effect correctly.

Figure \ref{fig3}{\bf a} shows the temperature dependence of the variation of the stray field by mechanical rotation at $\Omega=\pm1.5$ kHz. 
The $\Delta B(+\Omega)$ decreases with decreasing temperature and disappears at 240 K.
Although the magnetization induced by magnetic field persists at 240 K, which is much higher than the magnetic compensation temperature ($T_{\rm M}=135$ K) as shown in the upper panel of Fig.\ref{fig3}{\bf b}, the magnetization in rotation is zero.
The result indicates that the net angular momentum disappears at 240 K: the angular momentum compensation point.
Below 240 K, surprisingly, the sign of $\Delta B(+\Omega)$ reverses, and the magnitude of the field increases toward 200 K. 
This means that the direction of the magnetic moment with respect to the angular momentum turns toward the opposite direction below 240 K, that is, gyromagnetic reversal occurs.

The magnetization induced by the rotation ($M_\Omega$) can be estimated from $\Delta B$ by using a dipole field model\cite{Ono2015,Ogata2017A,Ogata2017B}.
The lower panel of Fig.\ref{fig3}{\bf b} shows the temperature dependence of $M_\Omega$.
At $T_{\rm M}=$135 K, $M_\Omega$ becomes zero.
Although the net angular momentum is non-zero at $T_{\rm M}$, there is no response to the rotation because the spontaneous magnetization disappears.
At 135 K, the sign of $M_\Omega$ reverses, and below 135 K, the net magnetic moment and the net angular momentum become antiparallel again.
Thus, the gyromagnetic reversal state is realized at temperatures between $T_{\rm M}$=135 K and $T_{\rm A}$=240 K.

\begin{figure}
\begin{center}
\includegraphics[width=0.6\linewidth]{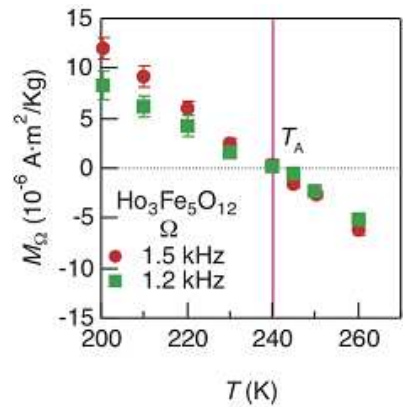}
\caption{ 
The magnetization of Ho$_3$Fe$_5$O$_{12}$ due to mechanical rotation ($M_\Omega$) at $\Omega=1.5$~kHz (red solid circles) and $\Omega=1.2$~kHz  (green squares).
\label{fig4}}
\end{center}
\end{figure}

The angular momentum compensation temperature obtained at the rotation frequency of 1.2 kHz is equivalent to that at 1.5 kHz, as shown in Fig.\ref{fig4}{\bf a}, indicating that $T_{\rm A}$ does not depend on the rotation frequency.
This result is consistent with the fact that the net angular momentum $J_{\rm net}$ is zero at $T_{\rm A}$.
The Barnett effect disappears and $M_\Omega$ is zero because spin-rotation coupling  $H_{\rm SR} = -J_{\rm net} \cdot \Omega$ is zero regardless $\Omega$.

The temperature controllable Barnett effect measurement makes it possible to determine $T_{\rm M}$ and $T_{\rm A}$ of bulk ferrimagnets.
These compensation points have attracted attention from the viewpoint of developing next-generation magnetic devices, such as spin valve induced by the spin torque\cite{Xin2006} and high-speed switching device\cite{Stanciu2007,Vahaplar2009,Radu2011}.
In addition, recently, domain wall mobility was found to be enhanced at $T_{\rm A}$ in the measurement of field-driven domain wall motion on the ferrimagnetic metal GdFeCo\cite{Kim2017}.
The technique proposed here open new opportunities in the search for suitable ferrimagnets for device, because the measurement can be performed on all bulk samples whether metal or insulator, without  micro fabrication technique. 
Not limited to ferrimagnets, the technique will be essential for gyromagnetic properties such as the contribution of orbital angular momentum in 3--5$d$, 4--5$f$ electron compounds.

\subsection{Sample preparation.}\label{sec:sample}
A powder sample of Ho$_3$Fe$_5$O$_{12}$ was prepared by means of a solid-state reaction.
Fe$_{2}$O$_{3}$(4N) and Ho$_{2}$O$_{3}$(3N) were mixed in a molar ratio of 5:3 in an agate mortar.
The mixture was pelletized and heated up to $1200\ {}^\circ\mathrm{C}$.

\subsection{Experimental setup of Barnett experiment at low temperature.}\label{sec:setup}
An air-driven rotor (produced by JEOL) was used and customized to rotate the samples in the forward and backward directions at frequencies up to 1.5~kHz.
To perform measurements below room temperature, air flow was replaced with cooled nitrogen gas flow.
A long copper tube wound into a coil was inserted into the cryostat as a heat exchanger, and it was cooled by liquid nitrogen sprayed into the cryostat.
The compressed nitrogen gas was cooled in the heat exchanger, and then the gas temperature was controlled by using the heater.   
Thereafter, the nitrogen gas was blown on to the rotor enclosing the sample.
The gas temperature was monitored by the thermometer adjacent to the rotor and the value was fed back to the output of the heater.

The pelletized sample of Ho$_3$Fe$_5$O$_{12}$ was crushed in powder, and then the powder was packed in a spindle capsule with a diameter of 6 mm and length of 16 mm.
The stray field from the sample was measured with a fluxgate magnetic sensor (Fluxmaster, Stefan Mayer Instruments, Dinslaken, Germany) placed near the rotor, as shown in Fig.\ref{fig2}{\bf b}.
The measurement was performed with positive and negative rotation as one set, and it was repeated 10 times or more to remove background noise.
The stray field $\Delta B$ was obtained from $\Delta B=[\Delta B(+\Omega)-\Delta B(-\Omega)]/2$. 
The magnetization induced by the Barnett effect $M_\Omega$ was evaluated using the following dipole model: $M_\Omega=-4\pi \mu_0(R^2+L^2/4)^{3/2}\Delta B/m$, where $R$, $L$, and $m$ are the sensor--sample distance, sample length, and sample mass, respectively.
This measurement system was placed inside a double magnetic shield to suppress fluctuations of the environmental magnetic field, for example, due to geomagnetism.

\begin{acknowledgments}
This work was financially supported by ERATO, JST,
a Grant-in-Aid for Scientific Research on Innovative Areas Nano Spin Conversion Science (26103005) from MEXT, Japan, 
a Grant-in-Aid for Scientific Research A (26247063) from MEXT, Japan, 
a Grant-in-Aid for Scientific Research B (16H04023) from MEXT, Japan, 
a Grant-in-Aid for Scientific Research C (15K05153, 16K06805) from MEXT, Japan, 
and a Grant-in-Aid for Young Scientists B (16K18353) from MEXT, Japan.
\end{acknowledgments}

\end{document}